**Infrared permittivity of the biaxial van der Waals semiconductor α-MoO₃ from near- and far-field correlative studies**

Gonzalo Álvarez-Pérez[1,2], Thomas G. Folland[3], Ion Errea[4,5], Javier Taboada-Gutiérrez[1,2], Jiahua Duan[1,2], Javier Martín-Sánchez[1,2], Ana I. F. Tresguerres-Mata[1], Joseph R. Matson[3], Andrei Bylinkin[6,7], Mingze He[3], Weiliang Ma[8], Qiaoliang Bao[8,9], José Ignacio Martín[1,2], Joshua D. Caldwell[3,*], Alexey Y. Nikitin[5,10,*] and Pablo Alonso-González[1,2,*].

[1] Department of Physics, University of Oviedo, Oviedo 33006, Spain.
[2] Center of Research on Nanomaterials and Nanotechnology, CINN (CSIC--Universidad de Oviedo), El Entrego 33940, Spain.
[3] Department of Mechanical Engineering, Vanderbilt University, Nashville 37235, TN, USA.
[4] Fisika Aplikatua 1 Saila, University of the Basque Country (UPV/EHU) and Centro de Física de Materiales (CSIC-UPV/EHU), Donostia/San Sebastián 20018, Spain.
[5] Donostia International Physics Center (DIPC), Donostia/San Sebastián 20018, Spain.
[6] CIC nanoGUNE, Donostia/San Sebastián 20018, Spain.
[7] Moscow Institute of Physics and Technology, Dolgoprudny 141700, Russia
[8] Department of Materials Science and Engineering, and ARC Centre of Excellence in Future Low-Energy Electronics Technologies (FLEET), Monash University, Clayton 3800, Victoria, Australia.
[9] Institute of Functional Nano and Soft Materials (FUNSOM), Jiangsu Key Laboratory for Carbon-based Functional Materials and Devices, Soochow University, Suzhou 215123, Jiangsu, China
[10] IKERBASQUE, Basque Foundation for Science, Bilbao 48013, Spain.

* Correspondence to: pabloalonso@uniovi.es, alexey@dipc.org, josh.caldwell@vanderbilt.edu

**The biaxial van der Waals semiconductor α-phase molybdenum trioxide (α-MoO₃) has recently received significant attention due to its ability to support highly anisotropic phonon polaritons (PhPs) —infrared (IR) light coupled to lattice vibrations in polar materials—, offering an unprecedented platform for controlling the flow of energy at the nanoscale. However, to fully exploit the extraordinary IR response of this material, an accurate dielectric function is required. Here, we report the accurate IR dielectric function of α-MoO₃ by modelling far-field polarized IR reflectance spectra acquired on a single thick flake of this material. Unique to our work, the far-field model is refined by contrasting the experimental dispersion and damping of PhPs, revealed by polariton interferometry using scattering-type scanning near-field optical microscopy (s-SNOM) on thin flakes of α-MoO₃, with analytical and transfer-matrix calculations, as well as full-wave simulations. Through these correlative efforts, exceptional quantitative agreement is attained to both far- and near-field properties for multiple flakes, thus providing strong verification of the accuracy of our model, while offering a novel approach to extracting dielectric functions of nanomaterials, usually too small or inhomogeneous for establishing accurate models only from standard far-field methods. In addition, by employing density functional theory (DFT), we provide insights into the various vibrational states dictating our dielectric function model and the intriguing optical properties of α-MoO₃.**

Polaritons —hybrid light–matter excitations— in hyperbolic media have opened up new avenues for studying exotic optical phenomena, such as non-reciprocal Purcell enhancement,[1] negative refraction,[2] non-reciprocal polariton guiding,[3] reconfigurable metaoptics,[4,5] topological transitions of in-plane anisotropic polaritons[6] and directional nanoscale energy collimation.[7,8] Excitingly, PhPs with in-plane hyperbolic and elliptic dispersion have been recently discovered in the biaxial van der Waals (vdW) semiconductor α-MoO₃ within a series of spectral bands —the so-called Reststrahlen bands (RBs)— throughout the long-wave infrared range[9-11] (LWIR). This discovery places α-MoO₃ at a privileged stage for studying and manipulating light at the nanoscale, beyond what has been achieved with other materials supporting PhPs,[12] such as silicon carbide (SiC)[13-15] or hexagonal boron nitride (hBN).[16-19] Yet, to accurately predict the IR response of α-MoO$_3$ and thus to enable predictive capabilities for advanced optical devices, it is imperative to develop an accurate dielectric function model. As such, prior work using α-MoO$_3$ has relied on a dielectric function that was estimated from previously reported optic phonon frequencies.[10,20,21] This approach (i) assumes the accuracy of these prior reports of the phonon frequencies, and (ii) cannot accurately extract the high- and low-frequency dielectric constants nor the phonon damping constants. These parameters are critical not only for building an accurate model, but for providing insights into how the crystal lattice dictates the optical response of the material. Here, we use model fitting of polarized far-field IR spectroscopy on single thick flakes of α-MoO$_3$ to overcome these limitations, identifying the phonon frequencies, high- and low-frequency dielectric constants and damping rates. Moreover, in an effort to refine this far-field approach and to compensate for any errors induced by the limited lateral scale of the flake (150 x 650 μm), which prevents conventional ellipsometry measurements, we compare near-field measurements of the damping and the dispersion of PhPs propagating on a thin flake of α-MoO$_3$ to full-wave numerical simulations, analytical dispersions and transfer-matrix calculations that use the initially extracted dielectric function as input. Exceptional quantitative agreement is simultaneously attained between the model and far- and near-field results through an iterative process combining both correlated efforts, dramatically improving the fit and thereby strongly verifying the accuracy and robustness of the resulting dielectric function. Additionally, we employ DFT to give insights into the vibrational modes dictating the oscillators on which this dielectric function model is based and into the intriguing optical properties of the material. Our work therefore reports an accurate and predictive model for the IR dielectric function of α-MoO$_3$, an emerging vdW material for nanophotonics,[22,23] while also offering an innovative approach to extracting dielectric functions of nanomaterials, where the use of traditional methods is challenging[24-26] or even not possible.

The exotic properties of PhPs in α-MoO$_3$ stem from their strong anisotropy, a consequence of the crystal structure, which is orthorhombic (Figure 1a). In it, layers formed by distorted α-MoO$_6$ octahedra are weakly bound by vdW forces and all three lattice constants (a, b and c) are different, as are the three principal values $\varepsilon_x(\omega)$, $\varepsilon_y(\omega)$ and $\varepsilon_z(\omega)$ —corresponding to the [100], [001] and [010] crystal directions,[27] respectively— of the dielectric tensor $\hat{\varepsilon}(\omega)$:

$$\hat{\varepsilon}(\omega) = \begin{pmatrix} \varepsilon_x(\omega) & 0 & 0 \\ 0 & \varepsilon_y(\omega) & 0 \\ 0 & 0 & \varepsilon_z(\omega) \end{pmatrix} \qquad (1)$$

Traditionally, the dielectric function of a material is extracted using far-field polarized reflection and transmission or ellipsometry measurements. In this work, to extract the three principal values $\varepsilon_x(\omega)$, $\varepsilon_y(\omega)$ and $\varepsilon_z(\omega)$ of α-MoO$_3$ in the mid- and far-IR —between ~9 and ~23 μm free-space wavelengths (1111 - 435 cm$^{-1}$)— we initially carry out Fourier transform infrared (FTIR) micro-spectroscopy. This technique can capture the optical response of relatively small single crystals across a broad spectral bandwidth, as required for accurate and unique determination of the dielectric function of α-MoO$_3$. The process for using FTIR micro-spectroscopy to determine the dielectric function involves taking accurate reflection and transmission spectra from a flake, and then using a Kramers-Krönig consistent model to fit to the experimental results. This model is typically chosen as it provides useful insights into the vibrational and/or electronic states of the material.

While in principle FTIR micro-spectroscopy can determine the optical properties of extremely small —down to 10 x 10 µm— flakes, in practice this limits the spectral range, accuracy and signal-to-noise ratio of spectra, in comparison to measurements on large films, due to the diffraction limit of light. As α-MoO3 is a biaxial crystal with optic phonons that occur in the LWIR (λ > 9.9 µm), we acquired polarized FTIR reflection (Figure 2) and transmission (Supporting Information) spectra from the same relatively large area of an α-MoO3 single flake (red square in Figure 1b), to ensure that good quality data were obtained across the full spectral range of interest, with the polarizer aligned along both in-plane principal axes (see Experimental Section for the detailed experimental procedure). The flake, with lateral dimensions of about 150 µm by 650 µm and a thickness of approximately 3 µm (Supporting Information), was exfoliated from bulk α-MoO3 crystals and then transferred onto an AgCl substrate (Figure 1b), which is highly transparent across the entire spectral range of interest (see dielectric function in the Supporting Information).

As a result of the strong in-plane anisotropy in the optical response of α-MoO3, the reflectance spectra along the two in-plane directions, [100] and [001], are clearly different (Figure 2). There are several high-reflectivity bands covering different frequency ranges, termed RBs, which are originated by the polar nature of the bonds in the so-called polar crystals, which include α-MoO3. These bands result from the breaking of the degeneracy between the longitudinal (LO) and transverse optical (TO) phonons, and the latter becoming IR active. Consequently, a strong absorption occurs at the TO phonon, and the real part of the frequency-dependent dielectric permittivity becomes negative within the RB, leading to the well-reported high reflectivity of polar materials within this spectral region.[28] Thus, we can use the reflectance data to approximate the position of the TO and LO phonons by the onset of the RB on the low- and high-frequency sides, respectively, providing the starting parameters for extraction of the IR dielectric function.

Although the permittivity of such an anisotropic crystal is a complicated function of frequency ω, in practice it can be approximated by model functions over the frequency range of interest. A widespread choice for polar dielectrics is the well-known Lorentz (or TO-LO) model for coupled oscillators.[29,30] According to the number of high-reflectivity bands in the reflectance spectra in Figure 2, we consider three oscillators for $\varepsilon_x$ and one oscillator for $\varepsilon_y$ and $\varepsilon_z$, respectively:

$$\varepsilon_x(\omega) = \varepsilon_x^\infty \left( \frac{\left(\omega_{x1}^{LO}\right)^2 - \omega^2 - i\gamma_{x1}\omega}{\left(\omega_{x1}^{TO}\right)^2 - \omega^2 - i\gamma_{x1}\omega} \right) \left( \frac{\left(\omega_{x2}^{LO}\right)^2 - \omega^2 - i\gamma_{x2}\omega}{\left(\omega_{x2}^{TO}\right)^2 - \omega^2 - i\gamma_{x2}\omega} \right) \left( \frac{\left(\omega_{x3}^{LO}\right)^2 - \omega^2 - i\gamma_{x3}\omega}{\left(\omega_{x3}^{TO}\right)^2 - \omega^2 - i\gamma_{x3}\omega} \right)$$

$$\varepsilon_y(\omega) = \varepsilon_y^\infty \left( \frac{\left(\omega_{y1}^{LO}\right)^2 - \omega^2 - i\gamma_{y1}\omega}{\left(\omega_{y1}^{TO}\right)^2 - \omega^2 - i\gamma_{y1}\omega} \right) \quad (2)$$

$$\varepsilon_z(\omega) = \varepsilon_z^\infty \left( \frac{\left(\omega_{z1}^{LO}\right)^2 - \omega^2 - i\gamma_{z1}\omega}{\left(\omega_{z1}^{TO}\right)^2 - \omega^2 - i\gamma_{z1}\omega} \right)$$

where $\varepsilon_i(\omega)$ denotes the i-th principal component of the permittivity tensor, and i = x, y, z stand for the three principal axes of the crystal, which correspond to the crystalline directions of α-MoO3, [100], [001], and [010], respectively. $\varepsilon_i^\infty$ represents the high-frequency dielectric constant, while $\omega_{ij}^{LO}$ and $\omega_{ij}^{TO}$ refer to the LO and TO phonon frequencies, along the i-th direction. $\gamma_{ij}$ represents the damping factor of the Lorentzian line shape that is derived from the phonon scattering rate and is inversely proportional to the phonon scattering lifetime. Finally, j is the subscript denoting the different phonon pairs along the same axis.

Using this model and our experimental data, we can in principle extract all the parameters that define the IR dielectric function in Equation (2) (Experimental Section), as previously employed for measuring the dielectric function of other anisotropic vdW materials, as hBN[18,31] or quartz.[32,33] However, this approach relies heavily upon reflection spectra collected at low angles, which is not very sensitive to absorption and thus damping, making the fitting procedure weakly sensitive to the phonon damping $\gamma_{ij}$. Furthermore, experimental non-idealities influence the fitted $\varepsilon_i^\infty$ and LO phonon energy values (Supporting Information). On the other hand, near-field polariton imaging by s-SNOM is extremely sensitive to both damping and LO phonon energies. More specifically, near the LO phonon frequency, the PhP wavelength is highly sensitive to small variations in the permittivity, while the measured propagation length is dictated by damping. As such, by extracting the PhP wavelength and propagation length from s-SNOM measurements on thin flakes at multiple incident laser frequencies, these parameters can be determined with dramatically improved accuracy. Even so, s-SNOM is not highly sensitive to other parameters, such as the high-frequency permittivity, and it is difficult to obtain measurements close to the TO phonon due to strong absorptive losses. Therefore, our procedure for extracting the permittivity of α-MoO$_3$ is based on an iterative effort utilizing the two correlated measurements in the far- and near-field, respectively.

Particularly, the free parameters, $\varepsilon_i^\infty$, $\omega_{ij}^{LO}$, $\omega_{ij}^{TO}$, and $\gamma_{ij}$, in Equation (2), are initially adjusted so that the simulated reflectance more accurately reproduces the experimental far-field FTIR spectra (Figure 2). To do so, the system is modeled as a three-layer structure (air/α-MoO$_3$/AgCl). After this first adjustment of the parameters (Supporting Information), they are provided as inputs to model near-field experimental data. For this purpose, we use the recently derived analytical dispersion of polaritons propagating in a biaxial slab embedded between two semi-infinite media:[34]

$$q = \frac{\rho}{k_0 d}\left[\arctan\left(\frac{\varepsilon_1 \rho}{\varepsilon_z}\right) + \arctan\left(\frac{\varepsilon_3 \rho}{\varepsilon_z}\right) + \pi l\right], \quad l = 0, 1, 2 \ldots, \quad (3)$$

where $q = k_t/k_0$ is the normalized in-plane momentum ($k_t^2 = k_x^2 + k_y^2$), $\varepsilon_1$ and $\varepsilon_3$ are the permittivities of the superstrate and substrate, respectively, d is the thickness of the α-MoO$_3$ flake, $k_0 = \omega/c$ is the wavevector in free space and $\rho = i\sqrt{\varepsilon_z/(\varepsilon_x \cos^2\alpha + \varepsilon_y \sin^2\alpha)}$, with α being the angle between the x axis and the in-plane component vector. On the other hand, an instructive way to visualize both the dispersion and the damping is via a false-color plot of the imaginary part of the Fresnel reflection coefficient, Im($r_p$), at real q and ω, obtained by means of transfer-matrix calculations.[35] Briefly, PhPs correspond to the divergences of the calculated reflectivity $r_p(q,\omega)$ of the anisotropic structure at complex momenta $q$. Both the analytical dispersion from Equation (3) and that inferred from transfer-matrix calculations are compared to the experimental dispersion (Figure 3), measured via polariton interferometry on an α-MoO$_3$ flake of thickness d = 120 nm (Supporting Information) placed on top of BaF$_2$, which, as AgCl, is highly transparent across the entire spectral range of interest (see dielectric function in Supporting Information), allowing for a finer tuning of the free parameters in Equation (2). Following this process, an excellent agreement between the experiment and our model is obtained. Remarkably, the ability to successfully fit the polaritonic response of a thin α-MoO$_3$ flake placed on a different substrate reinforces our claims of robust quantitative modeling of the far- and near-field IR response of this material, especially considering the strong thickness dependence of hyperbolic polaritons.[17]

To accurately adjust the damping and unambiguously verify that our dielectric function successfully accounts for the polaritonic response of α-MoO$_3$, we analyze the PhP propagation length from both simulated and experimental near-field images. To do so, we use our dielectric function model as input for full-wave numerical simulations (Figure 4b) that mimic s-SNOM measurements (Figure 4a) (see Experimental Section), carried out on a 120-nm-thick α-MoO$_3$ flake on top of BaF$_2$ at two representative frequencies: i) $\omega_0 = 910$ cm$^{-1}$ —within the lowest-frequency high-reflectivity band along the [100] crystal direction in Figure 2e (hyperbolic band[9,10] RB$_2$, 821 – 963 cm$^{-1}$); see Figure 4c and d— and ii) $\omega_0 = 990$ cm$^{-1}$ — within the high-reflectivity band along both in-plane directions in Figure 2e (elliptic band[9,10] RB$_3$, 957 – 1007 cm$^{-1}$); see Figure 4e and f. Within RB$_2$, we experimentally observe (Figure 4c) fringes along the [100] crystal direction, which indicate the excitation of in-plane hyperbolic PhPs in α-MoO$_3$, consistently with what has been previously reported[9,10]. By extracting a line profile from Figure 4c and fitting it to an exponentially-decaying oscillating function corrected by geometrical spreading factors (Supporting Information), we obtain a PhP wavelength of $\lambda_p^x = 820 \pm 25$ nm and a propagation length of $L_p^x = 1400 \pm 100$ nm. To account for these experimental results with our dielectric function, we adjust the parameter $\gamma_{x2}$ in the model — which determines the damping of PhPs in RB$_2$— and run a full-wave numerical simulation using the adjusted dielectric function. From the image resulting from this simulation (Figure 4d) we find the PhP wavelength and propagation length to be $\hat{\lambda}_p^x = 820 \pm 25$ nm and $\hat{L}_p^x = 1400 \pm 100$ nm, thus in excellent agreement with the experimental results. Within RB$_3$, we observe in the experimental image (Figure 4e) fringes along both in-plane directions, being the PhP wavelengths and propagation lengths $\lambda_p^x = 330 \pm 25$ nm and $L_p^x = 700 \pm 100$ mm, respectively along the [100] crystal direction, and $\lambda_p^y = 400 \pm 25$ nm and $L_p^y = 1100 \pm 100$ nm along the [001] direction. As before, to account for these experimental results with our dielectric function, we tune the parameter $\gamma_{z1}$ in our model —determining the damping of PhPs in RB$_3$— and run a full-wave numerical simulation using the adjusted dielectric function. From the resulting image (Figure 4f), we find the PhP wavelengths and propagation lengths to be $\hat{\lambda}_p^x = 325 \pm 25$ nm and $\hat{L}_p^x = 800 \pm 100$ nm along the [100] crystal direction and $\hat{\lambda}_p^y = 400 \pm 25$ nm and $\hat{L}_p^y = 1100 \pm 100$ nm along [001] direction. Again, we obtain excellent agreement with the experiment, unambiguously demonstrating the validity of our model.

As a further verification, we also carried out attenuated total reflectance (ATR) spectroscopy of an α-MoO$_3$ flake on top of BaF$_2$ (Supporting Information). FTIR-ATR has been proven to be an effective method for studying the properties of anisotropic 2D materials,[36] as it is sensitive to both dielectric and polaritonic resonant modes supported by these crystals. The good agreement (Supporting Information) between calculated and simulated spectra (using our dielectric function) again demonstrates the validity of our model and its broad applicability for predictive simulations.

We show in Table 1 the resulting frequencies and damping rates of the phonons of our model along all three principal axes and plot the extracted dielectric function in Figure 5 (middle panel). The parameters provided in Table 1 are closely related with prior experimental results[7-9] and previously reported phonon frequencies estimations[20,21] (Supporting Information). However, unlike earlier results, we identified two additional weak phonon bands aligned along the x axis ([100] crystal direction), and accurately determined the high-frequency dielectric permittivity along both the [100] and [001] directions. We note that the weak phonon spectrally located at 998 cm$^{-1}$ results in a discontinuity in the hyperbolic polariton dispersion relation (Figure 3), which has not been predicted using earlier experimental or theoretical models. We also note that the apparent strong damping of the optic phonon located near 506 cm$^{-1}$ is likely an artifact resulting from being spectrally near the cut-off of the FTIR mercury cadmium telluride (MCT) detector. As expected for a biaxial crystal, $\alpha$-MoO$_3$ exhibits different permittivities along the three principal axes throughout the entire spectral range studied ($\varepsilon_x \neq \varepsilon_y \neq \varepsilon_z$). Furthermore, over the spectral range extending from 545 to 1006 cm$^{-1}$, there is always at least one component $i$ exhibiting a negative Re($\varepsilon_i$) value, thus the crystal can support PhPs over this entire frequency range, having multiple hyperbolic regimes, each of them when at least one component $i$ fulfills Re($\varepsilon_i$) < 0.

To identify the vibrational character of the multiple optic phonons supported by $\alpha$-MoO$_3$, and to further test the validity of our dielectric function model, we have also calculated the permittivity tensor and vibrational eigenmodes using DFT (Supporting Information). The resulting *ab-initio* permittivity function qualitatively agrees with the experimentally extracted dielectric function (Figure 5, lower panel). This underlines the potential of employing first-principles methods to characterize the IR permittivity of emerging materials for nanophotonics. However, slight spectral shifts in the phonon frequencies are observed in the calculations with respect to the experiment, which is typical for DFT with semi-local exchange-correlation functionals for semiconducting oxides.[37] A further discrepancy is that, in the *ab-initio*-calculated dielectric function, RB$_2$ and RB$_3$ (Figure 5) do not spectrally overlap due to small underestimation and overestimation of the phonon frequencies along the [100] and [010] directions, respectively, in contrast with the experimental results.[9,10] We highlight that the *ab-initio* model also predicts several phonon modes beyond the spectral range of our experimental measurements, extending into the THz (Supporting Information). The validity of these predicted modes was verified through experimentally measured far-IR transmission spectra collected from $\alpha$-MoO$_3$ powder using a Nb-superconducting bolometer (Supporting Information). While we are not able to determine the exact frequencies for the phonons in this spectral band, the combination of the *ab-initio* model and THz transmission spectra suggest that this might be an interesting regime for future work.

In summary, we have extracted the full IR complex dielectric function of the biaxial vdW semiconductor $\alpha$-MoO$_3$ along all the three crystallographic directions. Its robustness is demonstrated by successfully numerically reproducing three different experimental measurements of flakes with significantly different thicknesses on different substrates. As the index of refraction of the substrate has been recently demonstrated to play a significant role influencing hyperbolic polariton propagation,[38] this further demonstrates the general nature of our reported dielectric function model. We achieved this by combining the virtues of far-field (broadband characterization and accurate determination of TO phonon energies) with near-field imaging (sensitivity to phonon damping and LO phonon energies) and advanced theoretical and numerical approaches, providing an alternative, more accurate approach to predicting the IR dielectric function —and thus the optical response— of 2D, nano and low-dimensional materials. Using the extracted permittivity, we envision that future experiments on IR nanophotonic and optoelectronic devices based on $\alpha$-MoO$_3$ will be streamlined and optimized using this more accurate model. Based on the breadth of exotic phenomena this material exhibits, such as the first reports of in-plane hyperbolicity within a natural, low-loss crystal,[9-

[11] and perspectives on negative refraction[2] or hyperlensing[39,40] and hyperfocusing,[39] the potential for unprecedented planar nanophotonic technologies is envisioned. Building on prior efforts utilizing local changes in the dielectric environment,[41] advanced concepts for reconfigurable planar metaoptics can be realized.[5,41] More broadly, structuring biaxial materials may offer new opportunities for making both passive and light emitting structures with unusual polarization states[8] —as in the creation of circularly polarized light. Furthermore, it may open new regimes for highly anisotropic dielectric resonators within the highly dispersive, extreme permittivity regime at frequencies below the TO phonon.[14,42,43]

**Experimental Section**

*Fourier-Transform Infrared Spectroscopy:* Mid-infrared reflectance measurements were undertaken using a Bruker Hyperion 2000 microscope coupled to a Bruker Vertex70v FTIR spectrometer, equipped with a broadband MCT detector (400 $cm^{-1}$ - 8000 $cm^{-1}$), and a wide range FIR beam-splitter (30 - 6000 $cm^{-1}$). We obtained off-normal (×36 Cassegrain, 25° average incidence angle) polarized reflection and transmission incidence spectra from the crystals. We used both KRS5 and polyethylene wire grid polarizers in order to optimize the spectral throughput of the system at the relevant phonon frequencies. The spectra were collected with a 2 $cm^{-1}$ spectral resolution and spatial resolution defined by the internal adjustable aperture of the microscope that was set to the size of the specific crystal of interest. All measurements were performed in reference to a gold film.

For FTIR-ATR micro-spectroscopy we used a modified Bruker optics ATR objective.[36] In brief, the objective is modified to reduce its incident angle, and azimuthal angular spread allowing us to produce s and p-polarized light at approximately 39° incidence angle. The Ge (100 μm diameter) prism is pressed into the substrate with a force of approximately 1N, allowing us to collect ATR spectra referenced to air. Spectra were taken with the crystal in two orientations with respect to the prism, allowing us to probe both crystal axes, under both polarization states of light.

In addition to the measurements on the α-$MoO_3$ crystals, the dielectric functions of AgCl and $BaF_2$ substrates were determined using unpolarized FTIR spectroscopy. FTIR reflectance and transmission spectra were collected at an angle of 40 degrees using a linear DLaTGs detector, under vacuum conditions. A broader description of such methods as they pertain to polaritonic materials can be found in a recent tutorial.[30]

*Fitting of Far-Field Experimental Data:* All dielectric function extraction was performed using the WVASE software (J.A. Woolam). Details on the obtained dielectric function are presented in the main text of the manuscript. Unlike infrared ellipsometry, there are a series of non-idealities which need to be accounted for when extracting the dielectric function from FTIR microscopy data.[30] First, we account for the incident angle and spread of the Cassegrain objective, (25° weighted average, 10 degree spread). Second, light from a Cassegrain objective is distributed over the full 180° azimuthal incidence spread, which means that s- and p-polarized light must be mixed during the fitting process. Finally, due to non-uniformity of the crystal surface, not all incident light will be collected, so a spectrally uniform linear scaling factor of 0.92 is used to change the percentage of collected light. Each of these three considerations were required to get an accurate fit of the model. Whilst ideally both reflection and transmission data would be considered for the fit, chromatic aberration induced by the dispersion of the substrate prevented the use of transmission data in our fit. Starting values for the fit were based off those from the *ab-initio* model, and the fit was performed accounting for results along both crystal axes at the same time. The fit was conducted by varying each parameter of the fit systematically, and once sufficiently close a global fit was performed to obtain the fully optimized parameters.

*Scattering-type Scanning Near-field Optical Microscopy (s-SNOM):* Infrared nano-imaging was performed using a commercially available s-SNOM (from Neaspec) where a metallized cantilevered atomic force microscope (AFM) tip is used as a scattering near-field probe. The tip oscillates vertically at the mechanical resonant frequency (around 270 kHz) of the cantilever, with an amplitude of about 50 nm. The tip is side-illuminated with p-polarized infrared light of frequency $\omega_0$ (from tunable $CO_2$ and quantum cascade lasers) and electric field **E**$_{inc}$. Acting as an infrared antenna, the Pt-coated tip concentrates the incident field into a nanoscale spot at the apex, which interacts with the sample surface and thus modifies the tip-scattered field **E**$_{sca}$. **E**$_{sca}$ is recorded with a pseudo-heterodyne Michelson interferometer. Demodulation of the interferometric detector signal at the nth harmonics of the tip oscillation frequency yields the complex-valued near-field signals $\sigma_n = s_n e^{i\phi_n}$, with $s_n$ being the near-field amplitude and $\phi_n$ being the near-field phase. By recording the near-field signals as a function of the lateral tip position, we obtain near-field images or line trace. In the case of probing a material supporting polaritons, the nanoscale 'hotspot' at the tip apex acts as a local source of polaritons. The tip-launched polaritons reflect at the flake edges and produce polariton interference, yielding fringes in the near-field images. The distance between the interference fringes corresponds to half the polariton wavelength, $\lambda_p/2$.

*Full-wave numerical simulations:* In s-SNOM experiments the tip acts as an optical antenna that converts the incident light into a strongly confined near field below the tip apex, providing the necessary momentum to excite PhPs. However, owing to the complex near-field interaction between the tip and the sample, numerical quantitative studies of s-SNOM experiments meet substantial difficulties in simulating near-field images.[44] To overcome these difficulties, we approximate the tip by a dipole source (with a constant dipole moment),[45] in contrast to the usual dipole model, in which the effective dipole moment is given by the product of the exciting electric field and the polarizability of a sphere.[46] We assume that the polarizability of the dipole is weakly affected by the PhPs excited in the α-MoO$_3$ flake, and their back-action onto the tip can be thus neglected. Calculating the amplitude of the near field, $|E_z|$, as a function of the dipole position (x,y), we simulate near-field images (we use COMSOL MULTIPHYSICS). The experimental s-SNOM images are well reproduced by our simulated images (see Figure 4c, d, e, f), which lets us conclude that the calculated field between the dipole and the α-MoO$_3$ flake, $E_z$, provides a valid numerical description of the signals measured by s-SNOM.

**Acknowledgements**

G.Á-P. and J.T.-G. acknowledge support through the Severo Ochoa Program from the I of the Principality of Asturias (grants No. PA-20-PF-BP19-053 and PA-18-PF-BP17-126, respectively). I.E. acknowledges support from the Spanish Ministry of Economy and Competitiveness (FIS2016-76617-P). J.M.-S. acknowledges support through a Clarín Marie Curie-COFUND grant from the Government of the Principality of Asturias and the EU (PA-18-ACB17-29), and the Ramón y Cajal Program (RYC2018-026196-I) from the Government of Spain. Q. B. acknowledges support from Australian Research Council (ARC, FT150100450, IH150100006 and CE170100039). J.D.C. was supported by the National Science Foundation (U.S.A.) under grant number U0048926. A.Y.N. acknowledges the Spanish Ministry of Science, Innovation and Universities (national project MAT2017-88358-C3-3-R) and Basque

Government (grant No. IT1164-19). P.A-G. acknowledges support from the European Research Council under Starting Grant 715496, 2DNANOPTICA.

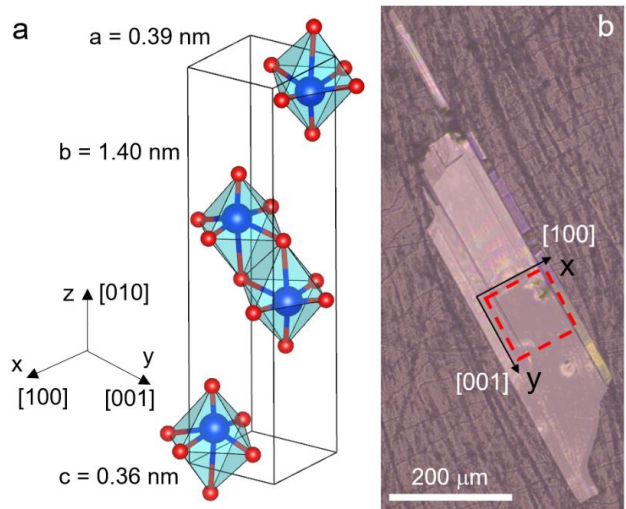

**Figure 1.** Crystalline structure of α-MoO$_3$. a. Sketch of the unit cell of α-MoO$_3$ and correspondence between the crystallographic directions [100], [001], [010] and the spatial coordinates x, y, z; the lattice constants are a = 0.396 nm, b = 1.385 nm and c = 0.369 nm. Blue/red spheres represent molybdenum/oxygen atoms. b. Optical image of an α-MoO$_3$ flake on top of AgCl. α-MoO$_3$ crystals typically appear to be rectangular owing to the anisotropic crystal structure. Labelled arrows indicate crystal directions.

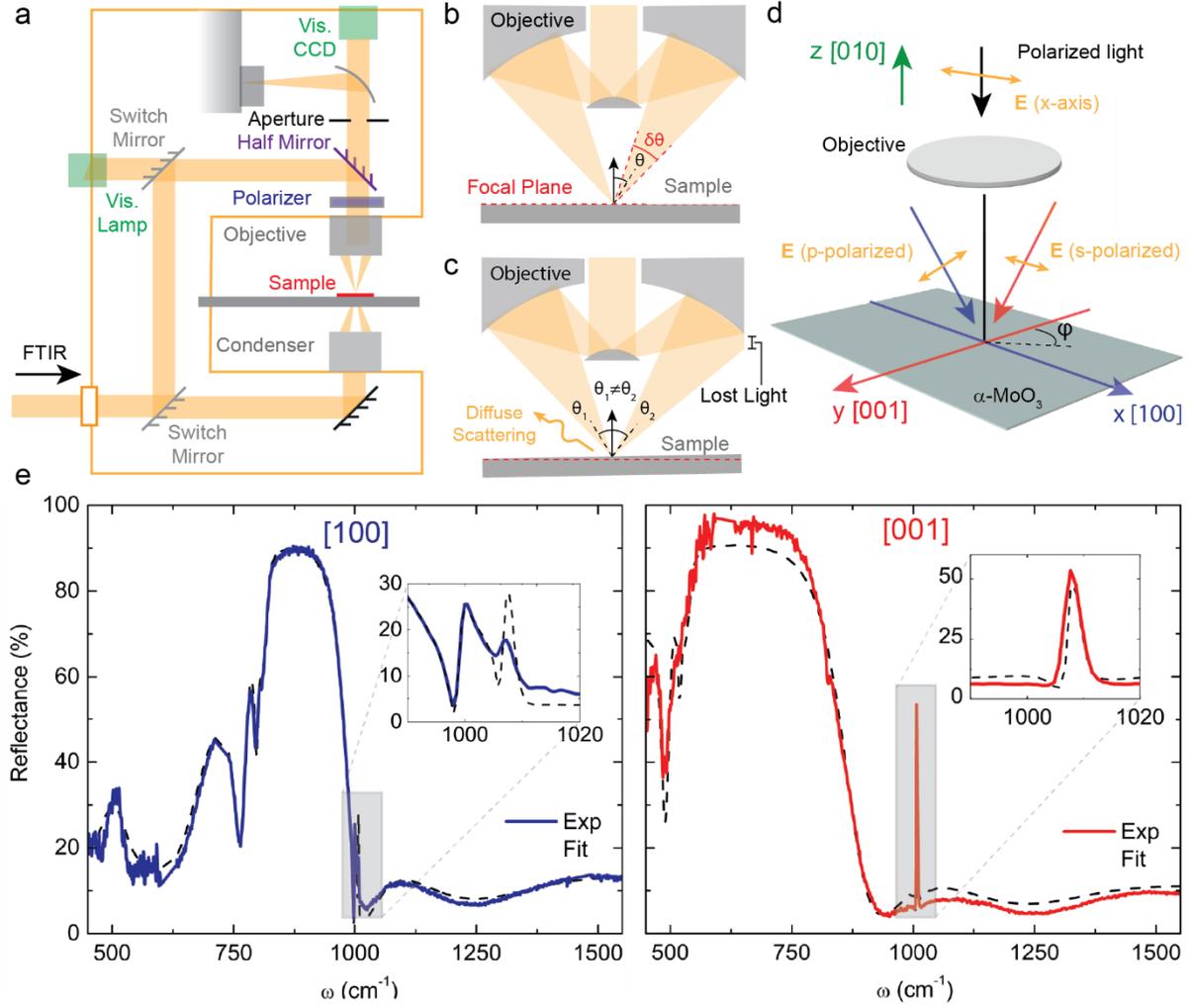

**Figure 2.** Far-field FTIR reflection measurements of the optical properties of an α-MoO$_3$ crystal slab (a) Schematics illustrating the setup for an FTIR microscope and beam path, adapted from ref. [30] and including (b) a Cassegrain objective with incidence angle θ and angular spread δθ, accounted for in our fitting process. In addition to angular spread, if the sample is not flat or oriented parallel with the focal plane, as in (c), then some of the light will not be collected through specular reflection outside the numerical aperture of the lens, or diffuse scattering. This is accounted for by fitting a spectrally flat scaling factor for the spectrum, which is close to 1 in our model. In addition to angular spread and imperfections in the specular reflection signal, the Cassegrain objective also produces a mixed polarization state of incoming light, as shown in (d). In this schematic, light is prepared propagating in the z axis towards the objective, and is polarized along the x axis. However, when it enters the objective it is reflected and re-distributed along different azimuthal incidence angles φ. The effect this has on the polarization state can be understood by considering the light propagating in the x-z and y-z planes, whilst the electric field is maintained in the x-direction. The light propagating in the x-z plane being incident with p-polarization, and light propagating in the y-z plane with s polarization (to ensure both have the electric field still oriented along the x axis). This mixing of s and p polarized light is accounted for in our model, and we fit the total combination of s- and p- polarized light with a value close to 1:1 mixing of s and p. (e) Far-field FTIR reflectance spectra of an α-MoO$_3$ thick crystal slab, with the polarization aligned along the [100] (top) and [001] (bottom) directions (shown in Fig. 1b). Dashed lines represent the initial fit used to extract

the α-MoO$_3$ dielectric function, then refined by near-field spectroscopy. Insets show the spectral region close to 1000 cm$^{-1}$ for clarity.

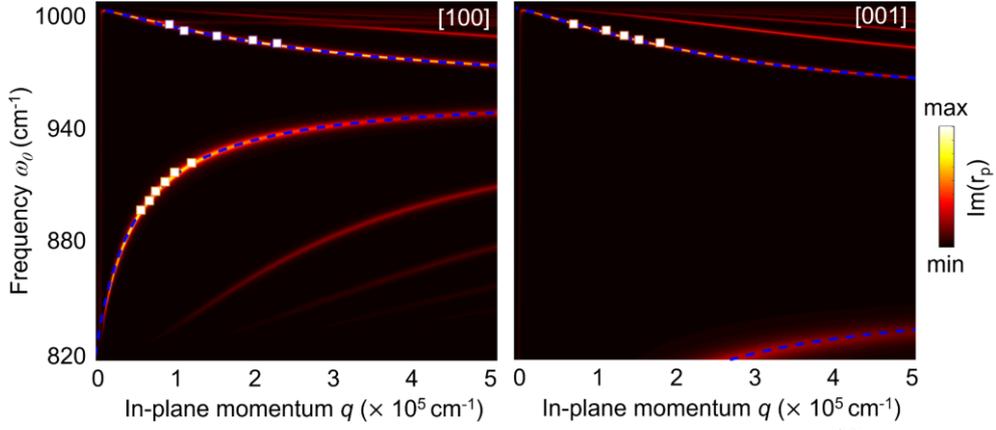

**Figure 3.** Dispersion of PhPs on a thin α-MoO$_3$ slab. Transfer-matrix[35] calculations (false color plot) and analytical dispersions[34] (dashed blue lines) of PhPs propagating on a 120-nm-exfoliated α-MoO$_3$ flake on top of BaF$_2$ along the [100] (left panel) and [001] (right panel) crystal directions. These calculations use the extracted permittivity from FTIR reflectance measurements as input. White dots indicate experimental data from monochromatic s-SNOM imaging.

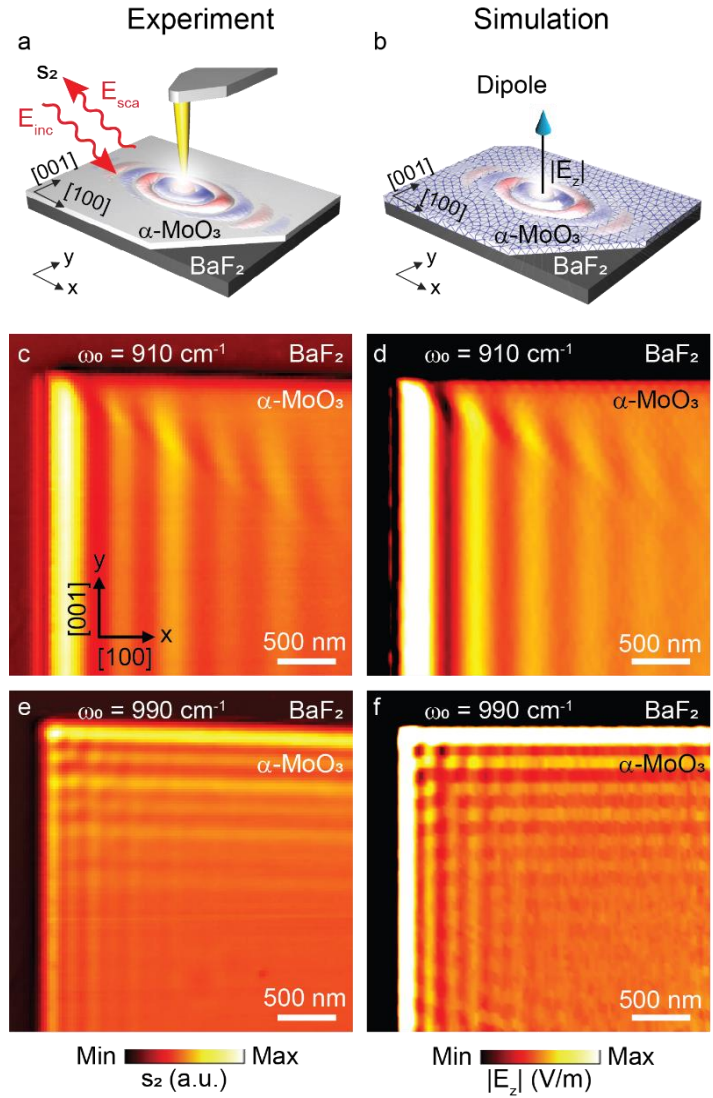

**Figure 4**. Near-field simulated and experimental images of in-plane anisotropic PhPs on an α-MoO$_3$ flake. a. Schematics of the s-SNOM experimental configuration and b. the full-wave numerical simulation model mimicking the experiment. c. (e.) Experimental near-field amplitude s$_2$ images and d. (f.) simulated z component of the electric field |E$_z$| of a 120-nm-thick α-MoO$_3$ flake on top of BaF$_2$ at incident frequency $\omega_0 = 910$ cm$^{-1}$ ($\omega_0 = 990$ cm$^{-1}$).

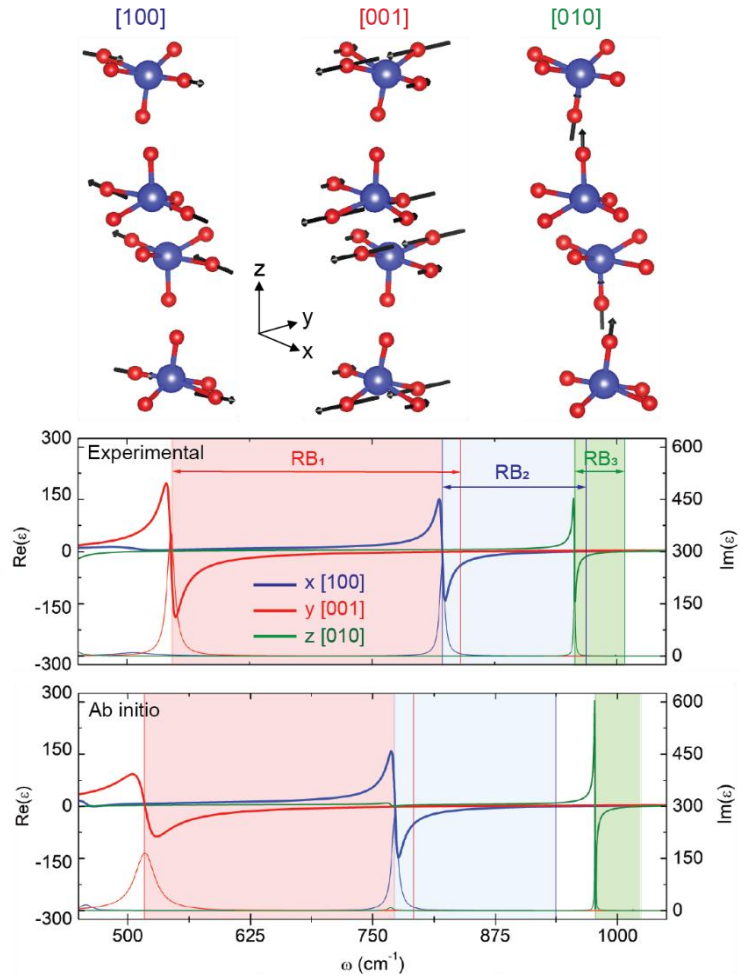

**Figure 5.** IR dielectric function of α-MoO$_3$. Experimental (middle panel) and *ab-initio* (lower panel) permittivity of α-MoO$_3$ along the three crystallographic axes. Shaded regions indicate the three Reststrahlen bands (RB$_1$, RB$_2$ and RB$_3$), corresponding to phonons on each of the crystallographic axes: [001], [100] and [010]. The lattice vibrations of these phonons are illustrated in the upper panel (molybdenum/oxygen atoms in blue/red, respectively).

**Table 1.** Parameters for the IR dielectric function of α-MoO$_3$, extracted from correlative far- and near-field experiments (left) and *ab-initio* calculations (right). The high-frequency permittivities are $\varepsilon_x^\infty$ = 5.78, $\varepsilon_y^\infty$ = 6.07 and $\varepsilon_z^\infty$ = 4.47 (fixed to the *ab-initio* value) for the experiment and $\varepsilon_x^\infty$ = 5.86, $\varepsilon_y^\infty$ = 6.59, $\varepsilon_z^\infty$ = 4.47 for the *ab-initio* calculation. The flake thickness used in the fit was 2.9 μm.

| Main axis | Mode index | Correlative far- and near-field experiments | | | *Ab-initio* calculation | | |
|---|---|---|---|---|---|---|---|
| i | j | $\omega_{ij}^{TO}$ (cm$^{-1}$) | $\omega_{ij}^{LO}$ (cm$^{-1}$) | $\gamma_{ij}$ (cm$^{-1}$) | $\omega_{ij}^{TO}$ (cm$^{-1}$) | $\omega_{ij}^{LO}$ (cm$^{-1}$) | $\gamma_{ij}$ (cm$^{-1}$) |
| x | 1 | 506.7 | 534.3 | 49.1 | 449 | 467 | 8.3 |
| x | 2 | 821.4 | 963.0 | 6.0 | 769 | 947 | 3.7 |
| x | 3 | 998.7 | 999.2 | 0.35 | 1016 | 1018 | 0.4 |
| y | 1 | 544.6 | 850.1 | 9.5 | 505 | 820 | 12 |
| z | - | - | - | - | 765 | 772 | 3.7 |
| z | 1 | 956.7 | 1006.9 | 1.5 | 976 | 1027 | 0.4 |

"-": not considered.

Supporting Information for

**Infrared permittivity of the biaxial van der Waals semiconductor α-phase molybdenum trioxide from near- and far-field correlative studies**

*Gonzalo Álvarez-Pérez, Thomas G. Folland, Ion Errea, Javier Taboada-Gutiérrez, Jiahua Duan, Javier Martín-Sánchez, Ana I. F. Tresguerres-Mata, Joseph R. Matson, Andrei Bylinkin, Mingze He, Weiliang Ma, Qiaoliang Bao, José Ignacio Martín, Joshua D. Caldwell,\* Alexey Y. Nikitin\* and Pablo Alonso-González.\**

**Contents**

**S1. AgCl dielectric function**

The infrared dielectric function of AgCl was extracted using the methods described in the main text. In brief, a 2-mm-thick AgCl window is mounted in the bench of a Fourier transform infrared (FTIR) spectrometer, and unpolarized reflection and transmission data are taken using a room temperature pyroelectric DLaTGs detector into the far infrared (Figure S1a). These data are fit using JA Wollams VASE software, using a phenomenological model with multiple Gaussian oscillators of the functional form

$$\varepsilon(E) = \varepsilon_\infty + \sum_{n=1}^{3} \varepsilon_1^n + i\varepsilon_2^n, \quad (S1)$$

$$\varepsilon_2^n(E) = A_n e^{-\left(\frac{E-E_n}{\sigma_n}\right)^2} - A_n e^{-\left(\frac{E+E_n}{\sigma_n}\right)^2},$$

$$\varepsilon_1^n(E) = \frac{2}{\pi} P \int_0^\infty \frac{\xi \varepsilon_2^n(\xi)}{\xi^2 - E^2} d\xi,$$

$$\sigma = \frac{Br_n}{2\sqrt{\ln(2)}},$$

where $A_n$ is the amplitude, $Br_n$ is the full width at half maximum (FWHM) of the gaussian, E is the frequency, $E_n$ is the center frequency and P is the Cauchy principal value. This model has been deliberately over-parameterized, and whilst it provides a good dielectric function fit, it does not provide physical information (such as phonon energies) about AgCl itself. The tabulated values for the fit are shown in Table S1. The dielectric function associated with this material is shown in Figure S1b, showing minimal dispersion in the region of interest for our experiments on α-MoO$_3$ samples.

| $\varepsilon_\infty = 4.89$ | A$_n$ | E$_n$ | Br$_n$ |
|---|---|---|---|
| Oscillator 1 | 1.14 | 12.3 | 266 |
| Oscillator 2 | 11.5 | 94.7 | 30.1 |
| Oscillator 3 | 69.0 | 13.3 | 153 |

**Table S1.** Fitting parameters for our AgCl dielectric function model.

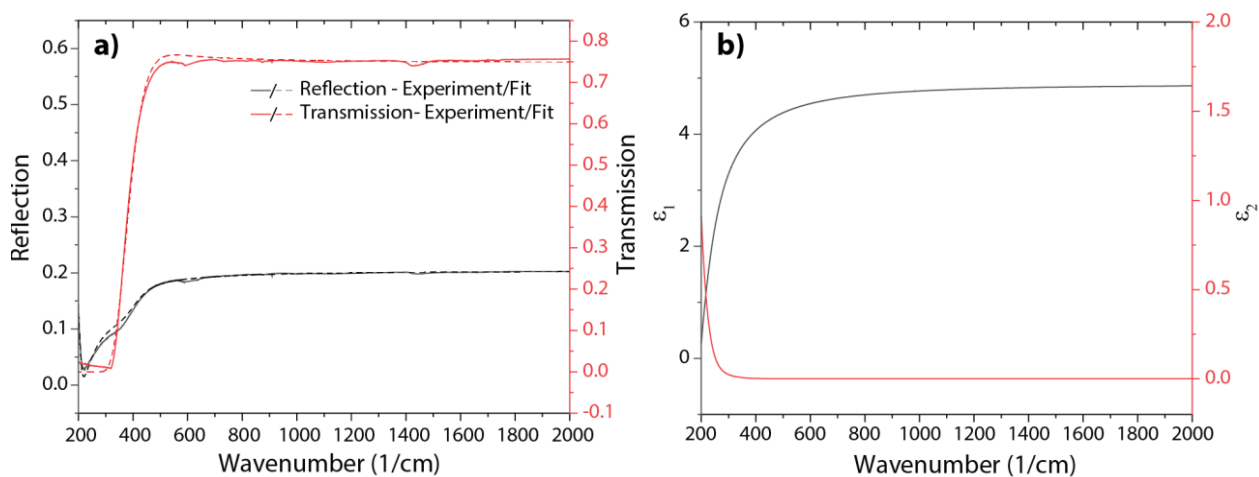

**Figure S1.** IR optical response (a) and fitted dielectric function (b) of AgCl.

**S2. Polarized FTIR transmission spectra.**

Transmission spectra were also taken through our sample of α-MoO$_3$, as shown in Figure S2. Whilst the spectral features of the transmission spectra are generally well reproduced, we note that the amplitudes of the transmission spectra are significantly offset. This is largely due to the collection of infrared light which is not transmitted through the crystal and prevents the use of these spectra for fitting. We also observe broad absorption bands between 1050 and 1150 cm$^{-1}$ which are not accounted for in our model. These are likely associated with weak multi-phonon processes in the α-MoO$_3$ flake.

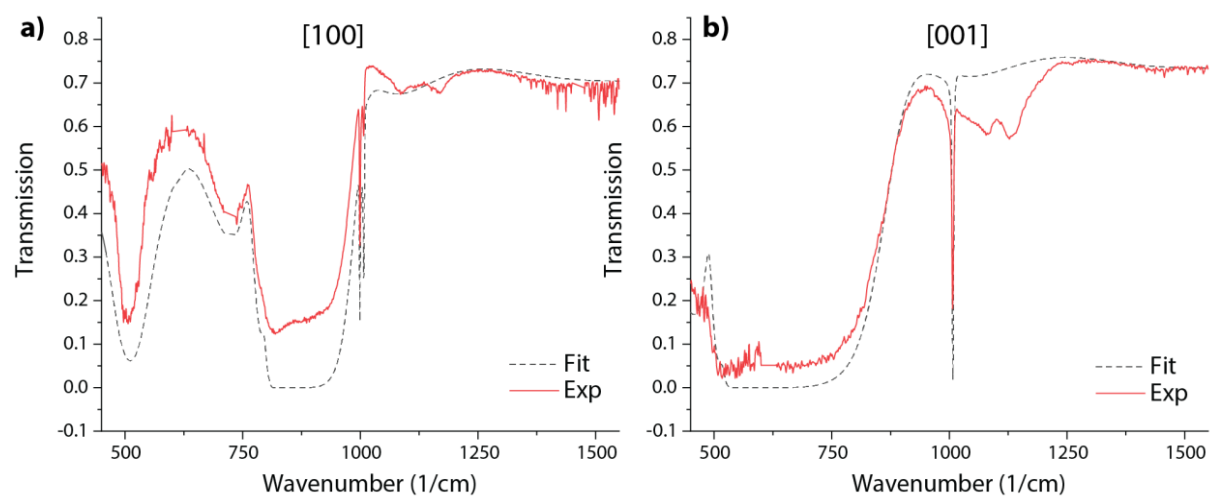

**Figure S2.** Transmission spectra along [100] or x axis (a), and along [001] or y axis (b).

## S3. Fit sensitivity of α-MoO₃ far-field data.

In this section we will assess fit 'uniqueness', to demonstrate the limitations of the far-field model —notably the limitations when it comes to phonon damping. A completely 'unique' parameter can be fit without coupling to any other parameters in the model, i.e. its value will be determined by some unique feature in the spectrum (for example a peak frequency). Fit uniqueness studies were performed by fixing the parameter under study at several values close to the global minimum, and then fitting all other parameters. We can then examine how much the mean squared error (MSE) increases relative to the global minimum value. Rapid increases in the MSE to small variations in the parameter value indicate a 'unique' and well fitted parameter, and vice versa. In this case we choose to analyze the parameters associated with the primary RB in the x direction: $\omega_{TO}$, $\omega_{LO}$, $\gamma$ and $\varepsilon_\infty$ (results shown in Figure S3).

The TO phonon clearly possesses the most unique fit value, with extremely strong increase in MSE away from the fitted frequency. The LO phonon, epsilon infinity and damping values show less dramatic variations in the MSE when compared to their global minimum value. In order to quantitatively compare all 4 parameters that fit our studies, we can use a figure of merit. We define the FOM as the change in the fit parameter when the MSE is increased by 1% of the global minimum, divided by the value at the global minimum. For the TO phonon we get 0.4%, for the LO phonon we get 0.6%, for epsilon infinity we get 4%, and for damping 96%. This demonstrates how our far-field fit is extremely weakly sensitive to the phonon damping value. We also note that the epsilon infinity value is not as unique as the TO and LO phonon energies. This is because non-idealities introduced into the model (nominally the amount of reflection collected, as well as the precise thickness) have coupling for this parameter.

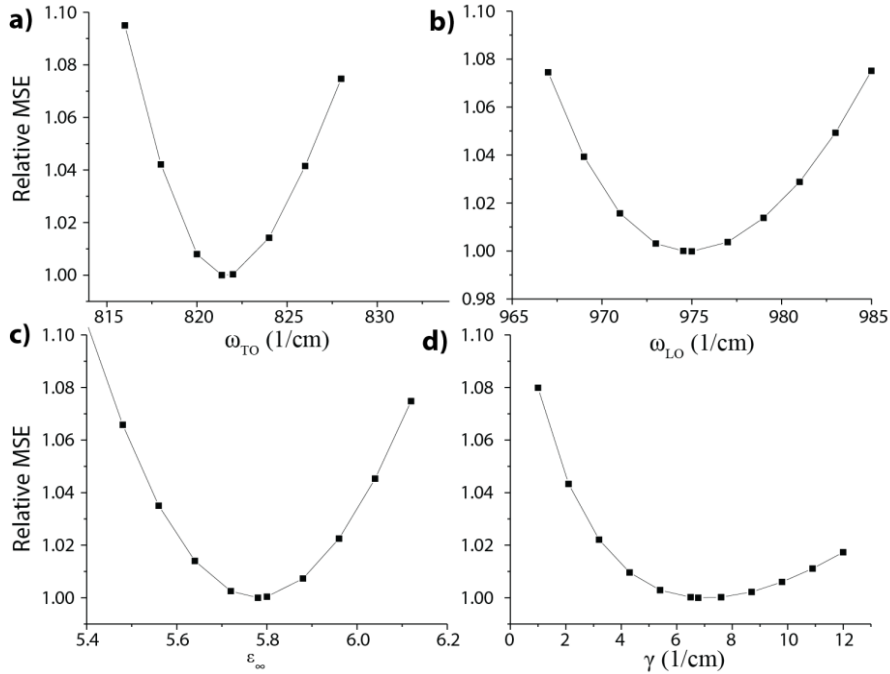

**Figure S3.** Fit uniqueness for the (a) TO phonon energy (b) LO phonon energy (c) epsilon infinity and (d) damping for the strong phonon band in [100] or x direction.

While the fit uniqueness value suggests that TO and LO energy values should be accurately determined from far-field data, coupling between parameters can complicate matters. Here we consider the coupling between the epsilon infinity value (which has dependence on experimental non-idealities) and the TO and LO phonon energies (Figure S4). We see that the epsilon infinity value couples to the LO phonon energy. This means that any error in the determination in epsilon infinity introduced by the experimental non-idealities and any associated local minima will have a knock-on effect on the LO phonon energy. Most notably, any inaccuracy in the thickness of the flake, as well as the percentage of the 1$^{st}$ reflection collected, can influence the value of epsilon infinity. The data from s-SNOM allows us to jump out of local minima, and find the best, and most realistic fit parameters.

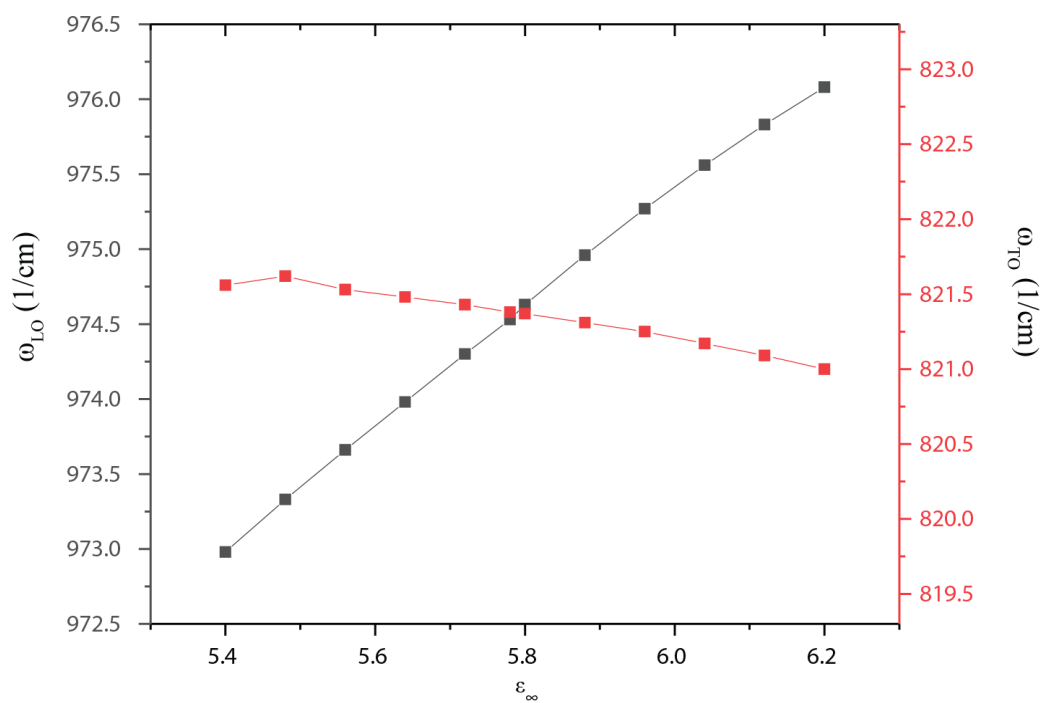

**Figure S4.** Coupling between epsilon infinity and LO phonon values.

### S4. Thickness and geometry of α-MoO$_3$ flakes

To extract material parameters with high accuracy, the geometry of the sample can be critical. In particular, the thickness of the flake has great influence both on the fitting of the FTIR spectra and in the PhP wavelength. The thicknesses of the α-MoO$_3$ flakes used throughout this work were determined using standard methods. On the one hand, the thickness of the flake employed in our far-field FTIR characterization was measured with a contact profilometer, while that of the thinner flake, used in near-field s-SNOM imaging, was determined via atomic force microscopy (AFM).

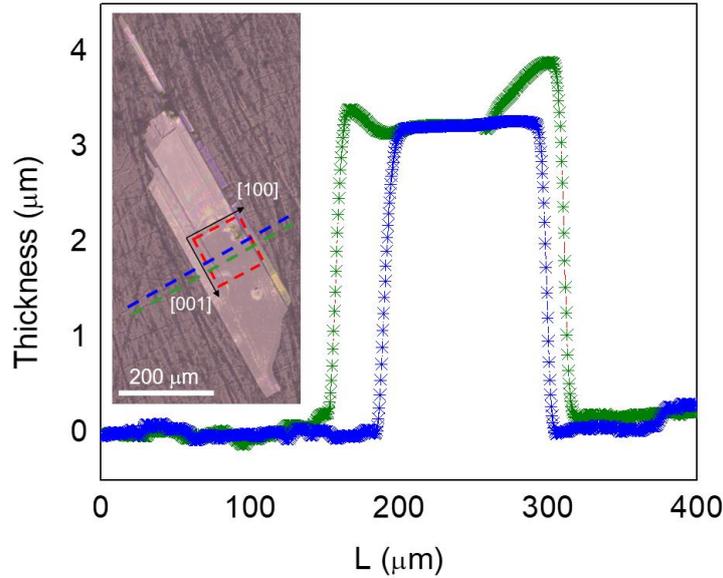

**Figure S5.** Determination of the thickness of the α-MoO$_3$ flake used for FTIR spectroscopy. The measurement was done using a contact profilometer, along the lines depicted in blue and green in the inset. The red square shows the area in which the FTIR spectra were acquired.

The thickness of the α-MoO$_3$ flake in which the FTIR spectra were acquired was determined using a contact profilometer (Figure S5), yielding a value of 3 ± 0.2 µm. Even though profilometers can be very accurate, we have considered an error of 200 nm bearing in mind i) the pressure exerted by the stylus onto the AgCl substrate, which is very soft, and ii) that the thickness is not perfectly constant in the area in which the FTIR spectra were acquired.

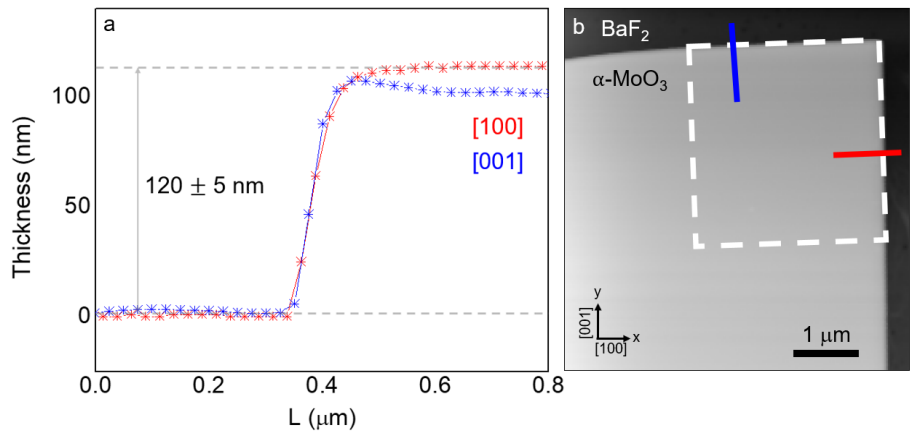

**Figure S6.** α-MoO$_3$ flake used for s-SNOM imaging. **a.** AFM profile along the two orthogonal in-plane directions: the flake thickness is 120 ± 5 nm. **b.** AFM image of the 90-degree-like corner used for near-field imaging, The white square shows the area in which s-SNOM imaging was performed.

Figure S6 shows an AFM image of the α–MoO$_3$ flake used for near-field imaging. Since the specific geometry of the edges of the flake can affect the near field distributions of PhPs, the corner in which we tested the predictive capabilities of our model was 90-degree-like. Averaging line profiles along both in plane directions, we obtain a flake thickness of 120 ± 5 nm, where the error stands for the thickness deviation along the flake edge and from data processing. Such profiles also show that the corners are steep and regular along the [010] or z direction, as needed for homogeneous launching of PhPs.

## S5. IR dielectric function of α-MoO₃, from far-field measurements

**Table S2.** Parameters for the IR dielectric function of α-MoO₃, extracted from far-field experiments. The high-frequency permittivities are $\varepsilon_x^\infty = 5.78$, $\varepsilon_y^\infty = 6.07$ and $\varepsilon_z^\infty = 4.47$ (fixed to the *ab-initio* value) for the experiment. The flake thickness used in the fit was 2.9 μm.

| Main axis i | Mode index j | $\omega_{ij}^{TO}$ (cm⁻¹) | $\omega_{ij}^{LO}$ (cm⁻¹) | $\gamma_{ij}$ (cm⁻¹) |
|---|---|---|---|---|
| x | 1 | 506.7 | 534.3 | 49.1 |
| x | 2 | 821.4 | 974.5 | 6.8 |
| x | 3 | 998.7 | 999.2 | 0.35 |
| y | 1 | 544.6 | 850.1 | 9.5 |
| z | 1 | 956.7 | 1006.9 | 0.65 |

## S6. BaF₂ dielectric function

The infrared dielectric function of BaF₂ was extracted using analogous methods as the ones described in Section S1 for extracting the dielectric function of AgCl: using a phenomenological model with multiple Gaussian oscillators of the functional form in Equation (S1). As that, this model for BaF₂ has been deliberately over-parameterized, and whilst it provides a good dielectric function fit, it does not provide physical information about BaF₂ itself. The tabulated values are provided in Table S3, while the infrared optical response and the fitted IR dielectric function of this material are shown in Figure S5, showing minimal dispersion in the region of interest for our experiments on α-MoO₃ samples.

| $\varepsilon_\infty = 2.68$ | $A_n$ | $E_n$ | $Br_n$ |
|---|---|---|---|
| Oscillator 1 | 1551.5 | 189.46 | 9.77 |
| Oscillator 2 | 2389.9 | 17.69 | 388.91 |
| Oscillator 3 | 0.1110 | 34.31 | 560.57 |

**Table S3.** Fitting parameters for our BaF₂ dielectric function model.

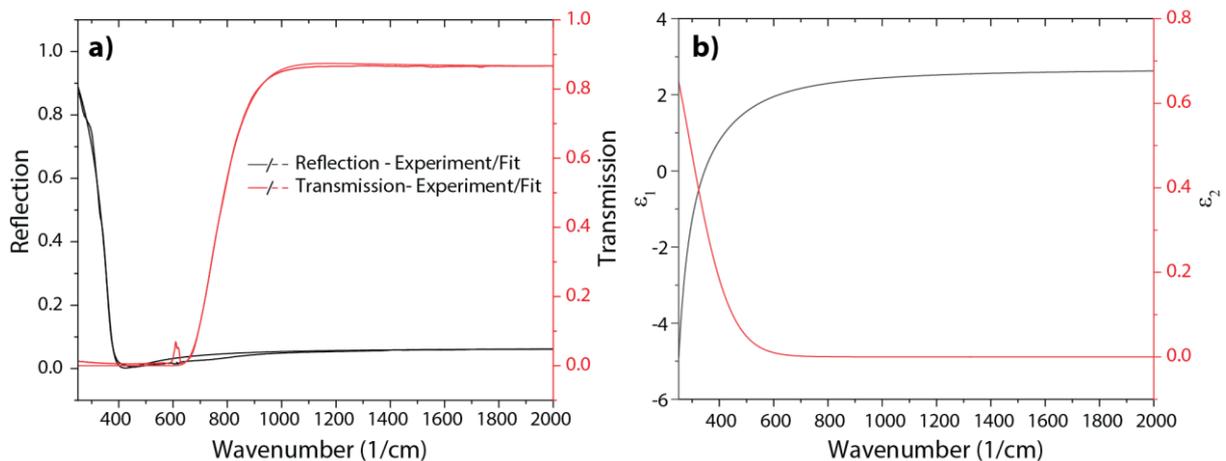

**Figure S7.** IR optical response (a) and fitted dielectric function (b) of BaF₂.

**S7. Propagation lengths and wavelengths of PhPs in α-MoO₃ from experimental s-SNOM profiles and full-wave numerical simulations**

The PhP propagation length $L_p$ and wavelength $\lambda_p$ is obtained by fitting line profiles of the polaritonic near field along the [100] and [001] crystal directions to oscillating signals with a factor accounting for dissipation described by an exponential decay of the PhPs field with distance x. The decay of polaritons away from an edge is due to a combination of damping ($\text{Im}(q_p) > 0$) and circular-wave geometrical spreading.[1,2] Thus, the oscillating signal of the polaritonic near field can be fitted with

$$\xi_{opt}(x) = \xi_0 + A\frac{e^{i2kx}}{\sqrt{x}} + B\frac{e^{ikx}}{x^a}, \qquad A, B, a > 0 \tag{S2}$$

where k is the complex PhP wavevector. The first term is the returning (to the s-SNOM tip) field for a damped circular wave reflected from a straight edge, with the PhP travelling a distance 2x. The second term interferes with the first, producing alternating fringe amplitudes. It arises because PhPs are not only generated/detected beneath the tip apex, but also at the edge of the flake travelling a distance x. Equation S2 can be easily transformed into

$$\xi_{opt}(x) = \xi_0 + A\frac{e^{-\frac{2x}{L_p}}\sin\left(4\pi\frac{x-x_A}{\lambda_p}\right)}{\sqrt{x}} + B\frac{e^{-\frac{x}{L_p}}\sin\left(2\pi\frac{x-x_B}{\lambda_p}\right)}{x^a}, \quad A, B, L_p, \lambda_p, a > 0, \tag{S3}$$

From the fitting procedure in Equation (S3) we directly extract the simulated and experimental PhP propagation length $L_p$ and wavelength $\lambda_p$. To fit the simulated profiles, we fix A = 0 and a = ½, as PhPs in our full-wave numerical simulations are launched by a point dipole (therefore there are no edge-lauched waves), and just $\lambda_p$-periodic waves with circular geometrical decay are present. On the other hand, in the experimental profiles, both contributions are present: in this case, since the geometrical decay of the PhP travelling the tip–edge distance only once is not known a priori, we allow for a variable decay[1] a ∼ 1. Figure S8 shows the profiles and the fittings performed to extract the polaritonic wavelength and propagation length, as well as theoretical values calculated from Equation (3) in the main text, found from $L_p = 1/\text{Im}(q)$ and $\lambda_p = 2\pi/q$, respectively. All experimental, simulated and analytical values are in excellent agreement.

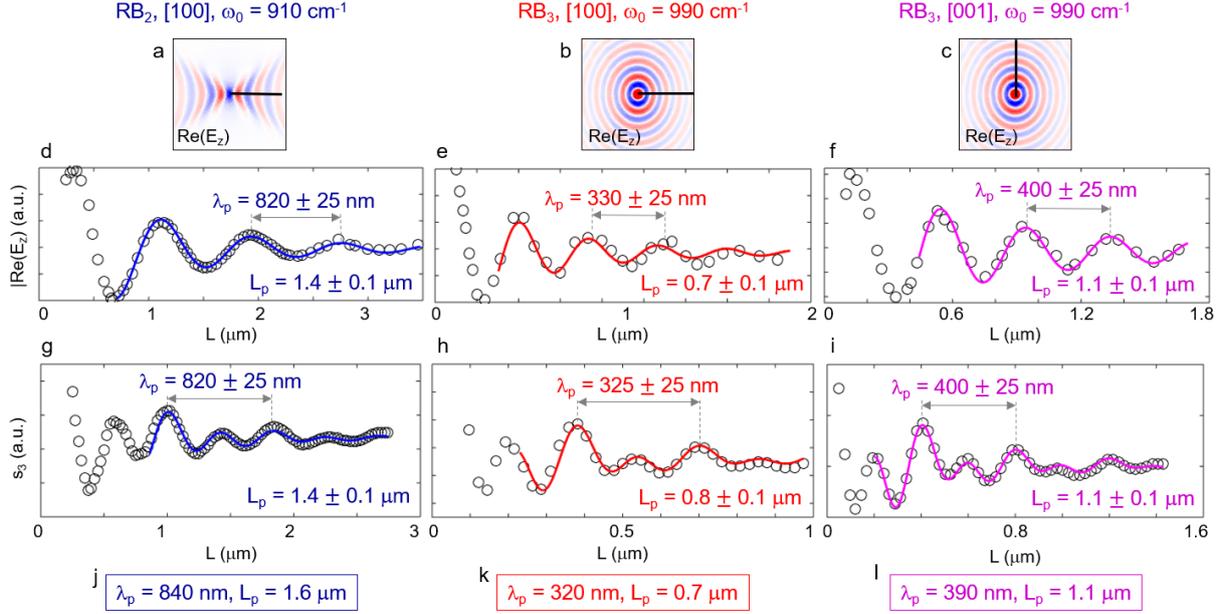

**Figure S8. a, b, c.** Field distribution of propagating PhPs launched by a point dipole oriented perpendicular to a 120-nm-thick α-MoO$_3$ layer for incident frequencies $\omega_0$ = 910 cm$^{-1}$ (in RB$_2$, left column) and $\omega_0$ = 990 cm$^{-1}$ (in RB$_3$, middle and right columns). **d, e, f.** Profiles extracted along the lines depicted in a, b, c (dots). By performing a fitting to Equation (S3) with A = 0 and a = ½,, (color lines), we find the PhP propagation lengths to be L$_p$ = 1.4 ± 0.1 μm (along [100] in RB$_2$), L$_p$ = 0.7 ± 0.1 μm (along [100] in RB$_3$) and L$_p$ = 1.1 ± 0.1 μm (along [001] in RB$_3$). All simulations were performed using our dielectric model as input. **g, h, i.** s-SNOM (amplitude) line traces (dots) along the 120-nm flake shown in Figures 4c,e of the main text. Fits to Equation (S3) setting a ∼ 1 are shown (color lines). The obtained propagation lengths L$_p$ = t$_0$ are L$_p$ = 1.4 ± 0.1 μm (along [100] in RB$_2$), L$_p$ = 0.8 ± 0.1 μm (along [100] in RB$_3$) and L$_p$ = 1.1 ± 0.1 μm (along [001] in RB$_3$). **j, k, i.** Theoretical values found from the analytical dispersion of PhPS, Equation (3) in the main text, yielding values in exceptional agreement with simulation and experiment: L$_p$ = 1.6 μm (along [100] in RB$_2$), L$_p$ = 0.7 μm (along [100] in RB$_3$) and L$_p$ = 1.1 μm (along [001] in RB$_3$).

## S8. Attenuated total reflectance spectra (ATR) of an α-MoO$_3$ flake on top of BaF$_2$

Finally, we examine the behavior of a α-MoO$_3$ crystal when probed using ATR microspectroscopy. This method is based on the Kretchmann-Raether configuration for launching hyperbolic polaritons[3,4]. In brief, a high index prism allows for coupling to modes with an effective index greater than one. One of the important parameters for this experiment is field orientation —as it is sensitive to both dielectric cavity resonances under s polarized light, and polariton resonances under p-polarized light. The optical anisotropy of α-MoO$_3$ implies we also need to consider the azimuthal incident angle due to the in-plane anisotropy. This means we collect four spectra —s and p polarized along the two different crystal axes (as shown in Figure S7). Within this experiment we are able to probe the polaritons launched in x and z axes, as we are limited by the optical transmittance of the Ge prism. The polariton/dielectric resonances manifest as absorption dips in the spectrum —with those located around 810 cm$^{-1}$ related to the [100] phonon, and those around 1000 cm$^{-1}$ associated with the [010] phonon. The simulations generally match our measured spectra, with the main differences being the presence of additional peaks in the green curves, and improved sharpness in the measured ATR spectra. This is most likely due to misalignment of the incident infrared beam with the crystal axes in the ATR measurement, and the overestimation of damping for our bulk dielectric function (as discussed in the main text). However, the measurement speaks to the usefulness of our dielectric function for predicting the behavior of α-MoO₃ samples.

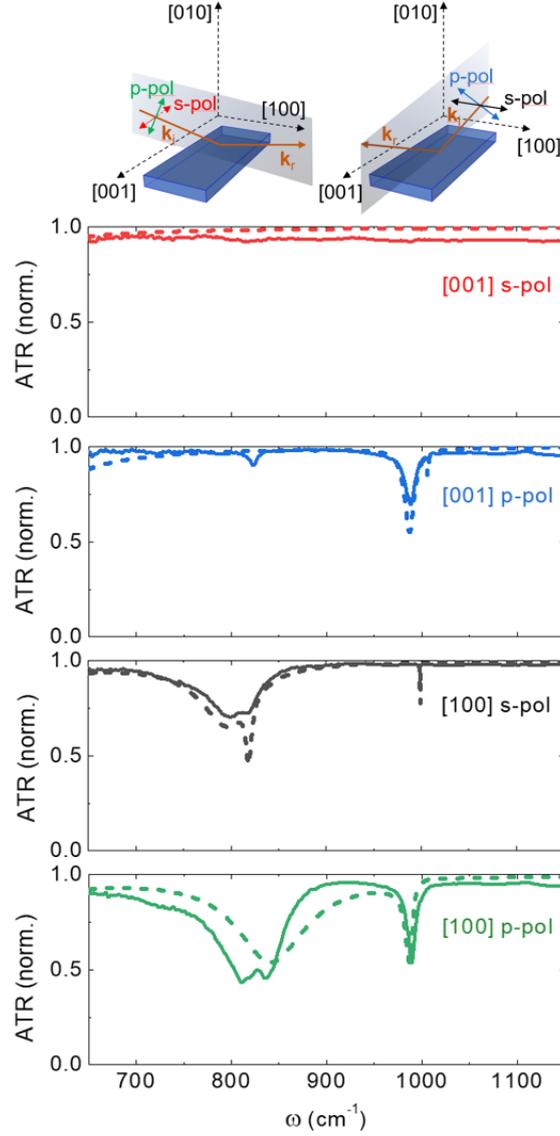

**Figure S9.** ATR-FTIR spectra (solid lines) on an α-MoO₃ flake (shown in Fig. 1b in the main text) along both in-plane crystal axes. Dashed lines indicate the simulated spectra, using our permittivity model as input.

**S9. Comparison of our IR dielectric function parameters with prior experimental results**

**Table S4.** Comparison of our extracted $\omega_{ij}^{TO}$, $\omega_{ij}^{LO}$, $\gamma_{ij}$ with values reported in prior works.

| Main axis | Mode index | This work | | |
|---|---|---|---|---|
| i | j | $\omega_{ij}^{TO}$ (cm⁻¹) | $\omega_{ij}^{LO}$ (cm⁻¹) | $\gamma_{ij}$ (cm⁻¹) |
| x | 1 | 506.7 | 534.3 | 49.1 |
| x | 2 | 821.4 | 963.0 | 6.0 |
| x | 3 | 998.7 | 999.2 | 0.35 |
| y | 1 | 544.6 | 850.1 | 9.5 |
| z | 1 | 956.7 | 1006.9 | 1.5 |

| Main axis | Mode index | Ref. [5] | | | Refs. [6,7] | | |
|---|---|---|---|---|---|---|---|
| i | j | $\omega_{ij}^{TO}$ (cm$^{-1}$) | $\omega_{ij}^{LO}$ (cm$^{-1}$) | $\gamma_{ij}$ (cm$^{-1}$) | $\omega_{ij}^{TO}$ (cm$^{-1}$) | $\omega_{ij}^{LO}$ (cm$^{-1}$) | $\gamma_{ij}$ (cm$^{-1}$) |
| x | 1 | - | - | - | - | - | - |
| x | 2 | 820 | 972 | 4 | 818 | 974 | - |
| x | 3 | - | - | - | - | - | - |
| y | 1 | 545 | 851 | 4 | 545 | 851 | - |
| z | 1 | 958 | 1004 | 2 | 962 | 1010 | - |

"-": not measured.

**Table S5.** Comparison of our extracted $\varepsilon_i^\infty$ with values reported in prior works.

| Main axis | $\varepsilon_i^\infty$ | | |
|---|---|---|---|
| i | This work | Ref. [5] | Refs. [6,7] |
| x | 5.78 | 4.0 | - |
| y | 6.07 | 5.2 | - |
| z | 4.47 | 2.4 | - |

"-": not measured.

## S10. Permittivity of α-MoO$_3$ from first principles

The dielectric function of α-MoO$_3$ is calculated fully *ab-initio* by making use of density functional theory (DFT) within the Perdew, Burke, and Ernzerhof parametrization of the exchange-correlation functional.[8] Ultra-soft pseudopotentials are employed in the calculations including 6 valence electrons for O and 14 valence electrons for Mo.

The dielectric tensor, which becomes diagonal in an orthorhombic crystal, is calculated within perturbation theory.[9] In this framework, the dielectric tensor for Cartesian directions α and β is given by

$$\varepsilon_{\alpha\beta}(\omega) = \varepsilon_{\alpha\beta}^\infty + 4\pi\chi_{\alpha\beta}(\omega), \tag{S4}$$

where $\varepsilon_{\alpha\beta}^\infty$ is the electronic contribution to the dielectric tensor (the high-frequency limit) and $\chi_{\alpha\beta}(\omega)$ is the susceptibility. The latter is expressed in atomic units as

$$\chi_{\alpha\beta}(\omega) = -\frac{1}{\Omega}\sum_i \frac{2\omega_i M_i^\alpha M_i^\beta}{\omega^2 - \omega_i^2 - 2\omega_i \Pi_i(\omega_i)}, \tag{S5}$$

where $\Omega$ is the unit cell volume, $\omega_i$ the phonon frequency of the i-th mode (the sum over i is restricted to phonon modes at the Γ point), $\Pi_i(\omega_i)$ the i-th phonon mode's self-energy due to phonon-phonon interaction taken at the phonon frequency, and $M_i^\alpha$ is the contribution of the i-th mode to the dipole moment:

$$M_i^\alpha = \sum_{\beta s} \frac{Z_s^{\alpha\beta} e_{is}^\beta}{\sqrt{2 M_s \omega_i}}. \tag{S6}$$

In Equation (S6), $Z_s^{\alpha\beta}$ is the effective-charge tensor for atom s, $M_s$ the mass of atom s, and $e_{is}^{\beta}$ the polarization vector of mode i for atom s along the Cartesian direction β.

Phonon frequencies, polarization vectors, effective charges, and the high-frequency limit of the dielectric function are calculated making use of density functional perturbation theory (DFPT)[10] as implemented in Quantum Espresso.[11,12] An 80 Ry energy cutoff was used for the plane-wave basis and a 800 Ry cutoff for the density. Electronic integrations were performed on an 8 x 2 x 8 grid. The phonon self-energy was calculated just including the so-called bubble contribution:[13,14]

$$\Pi_i(\omega) = -\frac{1}{2N_q}\sum_{qjk}|\Phi_{ijk}(0,q,-q)|^2 \left[\frac{2(\omega_{jq}+\omega_{k-q})[1+n_B(\omega_{jq})+n_B(\omega_{k-q})]}{(\omega_{jq}+\omega_{k-q})^2-(\omega+i\delta)^2} + \frac{2(\omega_{jq}-\omega_{k-q})[n_B(\omega_{k-q})-n_B(\omega_{jq})]}{(\omega_{k-q}-\omega_{jq})^2-(\omega+i\delta)^2}\right],$$
(S7)

In Equation (S7) $\omega_{jq}$ represents the j-th mode at the **q** point of the Brillouin zone, $N_q$ the number of **q** points in the sum, $n_B(\omega)$ is the Bose-Einstein occupation factor, δ is a small number (10 cm$^{-1}$ in our case), and $\Phi_{ijk}(0,\mathbf{q},-\mathbf{q})$ are the anharmonic third-order force constants transformed to the phonon mode basis.[14] The third-order anharmonic force-constants are calculated by finite-differences calculating atomic forces on displaced supercells created by the ShengBTE code.[15,16] The third-order force constants are calculated in a 2 x 1 x 2 supercell including interaction terms up to 5 nearest-neighbors. The self-energy is then calculated at 300 K by including in the sum over the **q** points a 16 x 4 x 16 phonon grid. The phonon frequencies and third-order force constants in this grid are obtained by Fourier interpolation. The dynamical matrices are calculated originally in a 6 x 2 x 6 **q** point grid. The *ab-initio* damping factors of the phonon mode $\omega_i$ reported in the main text correspond to the imaginary part of the phonon self-energy taken at the phonon frequency:

$$\gamma_i = -\mathrm{Im}\Pi_i(\omega_i).$$
(S8)

**S11. THz/far-IR transmission spectra collected from α-MoO₃ powder using a Nb-superconducting bolometer**

Whilst the small size of α-MoO₃ crystals precludes measurements in the far-IR, we can instead disperse a bulk powder across the surface of a THz/far-IR transparent material. Whilst this provides an ensemble averaged response of the powder, losing sensitivity to crystal axis, it does provide us with a rough estimate of the phonon energies, given by where the transmission/reflection is reduced/enhanced respectively. To perform these measurements, we dispersed α-MoO₃ crystals in IPA, via sonication for 30 min, and then drop cast on a piece of high density poly-ethylene (HDPE). We measure unpolarized reflection and transmission measurements at 40 degrees angle of incidence, shown in Figure S8. In general, the HDPE is weakly reflecting and moderately transmissive in the mid to far-IR. The powder sample generally shows reduced transmission and reflection —which is a consequence of scattering introduced by the crystals which have a size on the scale of the wavelength. However, at specific frequencies we observe a correlated dip in the transmission, and a peak in reflection. This suggests the presence of phonon modes at or close to these frequencies —notably 300, 349, 361 and 371 cm$^{-1}$. Whilst we cannot conclusively say that these correlated peaks and dips correspond to the exact phonon frequencies, this suggests that this frequency range could be of interest for future work.

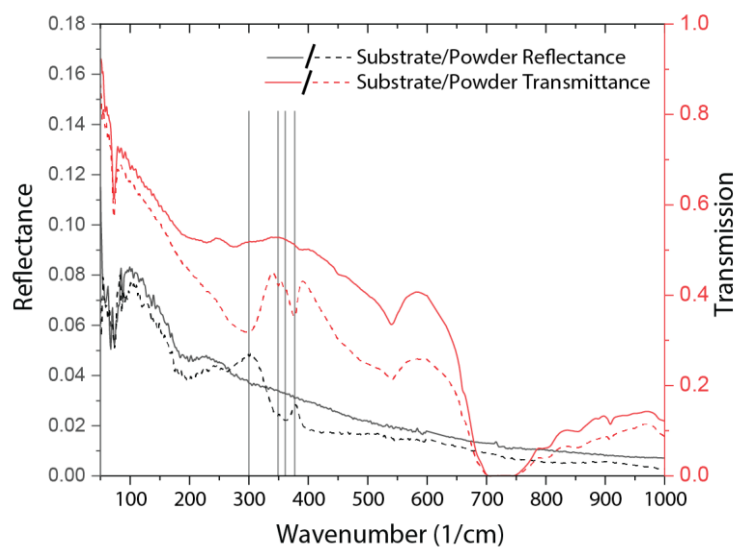

**Figure S10.** α-MoO$_3$ response in FIR/THz using reflection and transmission spectroscopy. Lines indicate correlated peaks/troughs (300, 349, 361 and 371 cm$^{-1}$), which will be associated with phonon modes.